\documentclass[fleqn,twoside]{article}
\usepackage{espcrc2}
\usepackage{amsmath}

\usepackage{graphicx}
\usepackage[figuresright]{rotating}

\title{Improving meson two-point functions by low-mode averaging}

\author{T. DeGrand\address[CU]{Department of Physics, University of Colorado,Boulder, CO 80309 USA}, S. Schaefer\addressmark[CU]\thanks{
    This work was supported by the US Department of Energy.
     Some simulations were performed on the Platinum IA-32 cluster at NCSA.}}
       
\begin{document}

\begin{abstract}
Some meson correlation functions have a large contribution
from the low lying eigenmodes of the Dirac operator. The contribution
of these eigenmodes can be averaged over all positions of the source.
This can improve the signal in these channels significantly. We test
the method for meson two-point functions.
\vspace{1pc}
\end{abstract}

\maketitle

\section{Introduction}

Simulations with light quarks face the problem of increasing noise
in meson two-point functions with decreasing quark masses.
Some of this noise is due to 
using only a single source for the meson instead of averaging over
all possible positions of this source.
We present  a method  to improve such meson 
correlation functions. More details can be found in \cite{DeGrand:2004qw};
for related work see \cite{Juge} and \cite{Giusti:2004yp}. 
The idea is to average the contribution
of the low-lying eigenmodes of the Dirac operator $D$ over
all positions of the source on the lattice. The contribution
of the high modes is calculated using the standard method, inverting
the Dirac operator on a single or a few quark
sources.
Ref.~\cite{Giusti:2004yp} has introduced the name low mode averaging (LMA)
for this method. LMA can have
a considerable impact on the noise in the two-point function
if it is  dominated by the low-lying modes. 
We expect it to be particularly effective at low quark masses and
for Dirac operators with good chiral symmetry like the 
overlap operator \cite{ref:neuberfer} which we use in our test.

\begin{figure}
\includegraphics[width=5cm,clip,angle=-90]{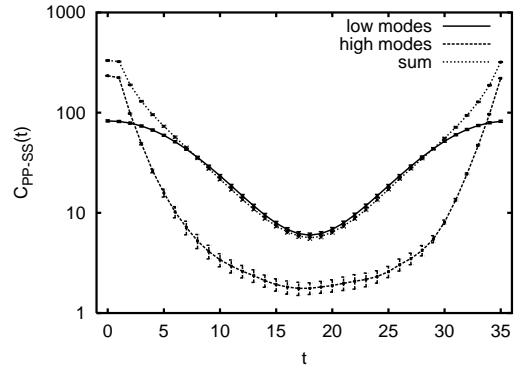}
\caption{\label{fig:1}The PP-SS two-point function. The contribution from
the $20$ lowest eigenmodes is shown.}
\end{figure}

An example for the dominance of the low modes
is the pseudo-scalar scalar difference shown
in Fig.~\ref{fig:1}. The contribution of the low lying modes
almost saturates the signal in the region $7 < t < 30$, from which 
the masses are to be extracted. 

For the simulation we have done, the improvement
comes essentially for free since the eigenmodes are used
to precondition the inversion of the Dirac operator. The 
associated gain in speed of the inversion alone justifies their computation.

\section{Method}
We are considering zero momentum two-point functions of the form
\begin{displaymath}
\begin{split}
\langle C(t)\rangle=\langle \frac{1}{L^3T}\sum_{x',x'',t''}  {\rm tr} &\Gamma_1G(x',t''+t;x'',t'') \times \\ & \Gamma_2
     G(x'',t'';x',t''+t) \rangle \ , 
\end{split}
\end{displaymath}
with $G$ the propagator matrix, 
\begin{displaymath}
(D+m)G(x',t';x,t)=\delta^3_{x,x'}\delta_{t,t'} \ , 
\end{displaymath}
and $\Gamma_i$ the Dirac matrix
corresponding to the meson in question. 

In a standard simulation, however, one does not compute the full propagator but only a
row, by inverting the Dirac operator on a localized source at, e.g., $(x',t')=(0,0)$. On
a given set of $N$ configurations one thereby
obtains an estimator $\langle C_1(t)\rangle$ for $\langle C(t)\rangle$. 
\begin{equation}
\begin{split}
\langle C_1(t)\rangle=\langle \frac{1}{L^3T}\sum_{x}  {\rm tr} &\Gamma_1G(0,0;x,t) \times \\ & \Gamma_2
     G(x,t;0,0) \rangle \ . \label{eq:2}
\end{split}
\end{equation}

However, using eigenmodes of the Dirac operator,
one can compute $C(t)$ itself. 
As this is in general not possible for all eigenmodes, we
split the propagator in a contribution from the low modes and one from the high modes
$G=G_L+G_H$ with $G_L$ the propagator in spectral representation
\begin{equation}
G_L(x,t;x',t')= \sum_{j=1}^n \frac{\langle x,t|j\rangle\langle j|x't'\rangle}{i\lambda_j+m}.
\end{equation}
The sum is over the $n$ lowest eigenmodes of the 
Dirac operator $|j\rangle$ with eigenvalues $i \lambda_j$.

The two-point function is thereby  separated into four different parts, one
from the low modes alone, one from the high modes and two interference terms
\begin{equation}
C(t)= C_{LL}(t)  + C_{HL}(t)  + C_{LH}(t)  + C_{HH}(t)  \ .
\end{equation}
The first contribution $C_{LL}$ can be expressed by the low-lying 
eigenmodes alone and can thus be averaged over all positions of the source.
The other contributions are restricted to the usual one or a few quark sources.

\section{Test of the method}
To test the method we use 80 quenched gauge configurations of size $12^3\times36$ generated 
with the Wilson gauge action at $\beta=5.9$. We use the overlap Dirac operator constructed
from HYP 
smeared gauge links. The low mode averaging is done using the 20 lowest eigenmodes
of the Dirac operator which are also used to precondition the computation
of the propagator. The bare  quark masses are between $0.015$ and $0.035$ which
correspond  to pseudo-scalar to vector meson mass ratios
 $m_{\rm PS}/m_{\rm V}$ ranging between about 0.4 to 0.64, see \cite{DeGrand:2003in} for
 further details.
  On each of the configurations the inversion
  of the Dirac operator was done on two Gaussian sources with radius $3a$,
  one located on time-slice $t=0$, the other on $t=16$. We average over these two positions.
  The $\Gamma$ matrices depend on the meson in question and we shall use the
  abbreviations given in Table~\ref{tab:t1}.

\begin{table}
\begin{center}
  \begin{tabular}{c|c|c|c|c}
  P& S& $A_\mu$&$ V_\mu$ &$B_{\mu\nu}$\\
  \hline
  $\gamma_5$ &$ 1$ &$ \gamma_5\gamma_\mu$ &$\gamma_\mu$ &$\gamma_\mu\gamma_\nu$
  \end{tabular}
  \end{center}
  \caption{\label{tab:t1}Labels for Dirac structure of currents.}
  \end{table}

\begin{figure}
\begin{center}
\includegraphics[width=5cm,clip,angle=-90]{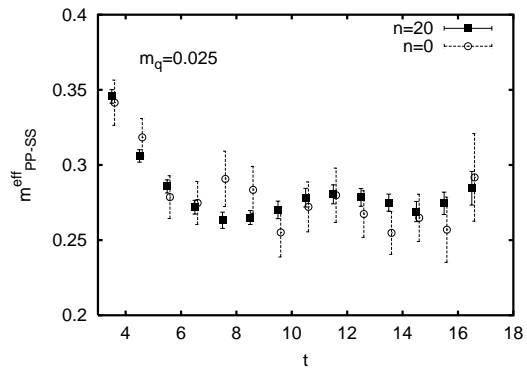}
\vspace{-0.5cm}
\end{center}
\caption{\label{fig:2} The effective mass plots for the $PP-SS$ correlator
at $m_q=0.025$. The open circles are the effective mass without
low mode averaging, the the signal given by the filled squares is
averaged using 20 eigenmodes.}
\end{figure}

In Fig.~\ref{fig:2} we plot the effective mass for the pseudo-scalar scalar difference
correlator. (Subtracting the scalar two-point function removes the contributions of the
zero modes.) Using LMA with 20 eigenmodes,
we observe a significant improvement in the noise of the signal.
In Ref.~\cite{DeGrand:2004qw} we show similar plots for other correlators.

\begin{figure}
\begin{center}
\includegraphics[width=4cm,clip,angle=-90]{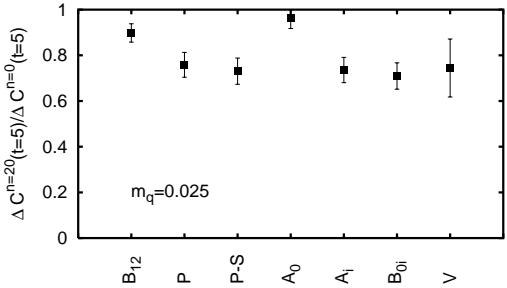}
\end{center}
\vspace{-0.5cm}
\caption{\label{fig:3} Ratio of the error bars for $n=20$ compared $n=0$ at
time-slice $t=5$ for the different meson correlators.
Data uses two source points for the high eigenmode part of the correlator.}
\end{figure}

To give an impression of the improvements in the various channels,
the ratio of the uncertainty at time-slice $t=5$ with LMA over 
20 eigenmodes to the uncertainty without LMA is shown in Fig.~\ref{fig:3}.
In the $\gamma_5$, $\gamma_5\gamma_i$, $\gamma_0\gamma_i$ and $\gamma_i$ channels,
we find an improvement of roughly $30\%$, which corresponds to
a factor of 2 in statistics. The $\gamma_5\gamma_0$ and $\gamma_i\gamma_j$ cannot
profit from the LMA. It is known that low modes make a small contribution to these
correlators \cite{DeGrand:2003sf}.

\begin{figure}
\begin{center}
\includegraphics[width=4.5cm,clip,angle=-90]{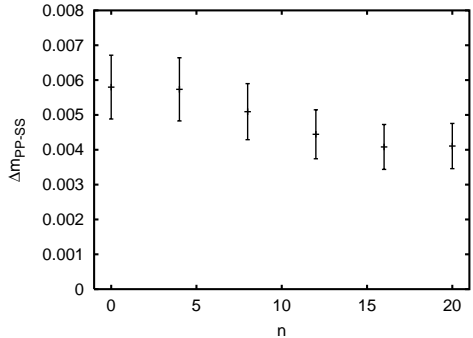}
\vspace{-0.5cm}
\end{center}
\caption{\label{fig:4} Dependence of the uncertainty in the  extracted $PP-SS$ 
mass on the number of
eigenmodes included at $a m_q=0.025$. %The fit range is from $t=8$ to 18.
}
\end{figure}

\begin{figure}
\begin{center}
\includegraphics[width=4.5cm,clip,angle=-90]{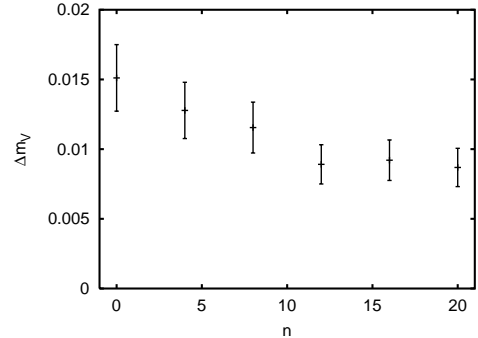}
\vspace{-0.5cm}
\end{center}
\caption{\label{fig:5} Same as Fig.~\ref{fig:4} for the $B_{0i}$ mass.
}
\end{figure}

The ultimate goal of this method is to improve the errors of the
extracted meson masses. In Figs.~\ref{fig:4} and \ref{fig:5}  we show the dependence
of the error of the masses on the number of eigenmodes used
in the LMA. We find that the improvement seems to saturate at 12 to 16
modes included and reaches about $30\%$ for the PP-SS mass and about
$40\%$ for the vector meson from the $B_{0i}$.

\section{Conclusion}
We tested a method to improve meson two-point functions in lattice QCD.
We found an improvement in the uncertainty of some masses corresponding
to a factor of 2 in statistics. This improvement comes at virtually
no additional cost because the cost of computing the low
eigenmodes is justified by the acceleration of the 
inversion of the Dirac operator alone.

\end{document}